\documentclass[preprint,aps,prb]{revtex4}
\usepackage[dvips]{graphicx}
\usepackage{color,array,dcolumn}
\usepackage{amsmath}
\usepackage{amssymb}

\begin{document}

\title{Including screening in van-der-Waals corrected DFT calculations. 
The case of atoms and small molecules physisorbed on graphene}

\author{Pier Luigi Silvestrelli}
\affiliation{Dipartimento di Fisica e Astronomia, 
Universit\`a di Padova, via Marzolo 8, I--35131, Padova, Italy,
and DEMOCRITOS National Simulation Center, of the Italian Istituto 
Officina dei Materiali (IOM) of the Italian National 
Research Council (CNR), Trieste, Italy}


\begin{abstract}
\date{\today}
The DFT/vdW-QHO-WF method, recently developed to include the van der  
Waals (vdW) interactions in approximated Density Functional Theory (DFT)
by combining the Quantum Harmonic Oscillator model with 
the Maximally Localized Wannier Function technique, is applied to the
cases of atoms and small molecules (X=Ar, CO, H$_2$, H$_2$O) weakly 
interacting with benzene and with the ideal planar graphene surface.
Comparison is also presented with the results obtained by other 
DFT vdW-corrected schemes, including PBE+D, vdW-DF, vdW-DF2, rVV10,
and by the simpler Local Density Approximation (LDA) and semilocal Generalized 
Gradient Approximation (GGA) approaches. 
While for the X-benzene systems all the considered vdW-corrected schemes
perform reasonably well, it turns out that an accurate description of the 
X-graphene interaction requires a proper treatment of many-body contributions
and of short-range screening effects, as demonstrated by 
adopting an improved version of the DFT/vdW-QHO-WF method. 
We also comment on the widespread attitude of relying on LDA 
to get a rough description of weakly interacting systems.
\end{abstract}

\maketitle

\section{Introduction}
Nowadays the importance of graphene both from a theoretical 
point of view and considering the ever growing 
interest for many possible nanotechnology applications cannot be
surely overemphasized.\cite{Novoselov,Vanin,Kim}
Of particular relevance is the investigation of isolated molecules
interacting with graphene. In fact, for instance, 
the operational principle of graphene devices is mainly based on
changes in their electrical conductivity due to gas molecules
adsorbed on the graphene surface.\cite{Novoselov} 
The study of the interactions of
graphene with water is also very important as a model for the characterization
of the interface between water and hydrophobic substrates,\cite{Rubes} and
considering the recently
reported application of graphene as an atomically flat coating for
atomic force microscopy used for
investigating the growth of water adlayers on a substrate.\cite{Heath}
In addition, a single water molecule adsorbed on graphene 
represents a weakly interacting system involving a complex
mixture of Hydrogen bonding, electrostatic, and van der Waals
interactions, thus providing a significant test model for any electronic
structure theory.\cite{ambrosetti}

Moreover, observations indicate that 
the cold regions of the interstellar medium contain clouds made of 
atoms, radicals and simple, ``astrobiological'' molecules such 
as H$_2$, CO, H$_2$O,.., as well as small solids typically 
composed of amorphous water, silicates and 
carbon grains in the form of graphite, amorphous structures and 
polycyclic aromatic hydrocarbon molecules, such as benzene.\cite{Lakhlif} 
More complex molecules are probably formed by surface-catalysed chemical 
reactions on low temperatures interstellar dust grains.
Clearly a pre-requisite for describing such processes is the possibility
of accurately reproducing the interactions between isolated small molecules
and grain-surface models.

Carbon-based nanomaterials have also
attracted much attention because of their suitability as
materials for gas storage. In particular, the reported high
hydrogen uptake of these materials make them attractive for hydrogen 
storage devices in fuel-cell-powered electric vehicles.\cite{Geim}
Among the available carbon nano-materials, the graphene sheet is the 
simplest one and may store hydrogen on both side of its structure.
The hydrogen-storage process implies the H atomic chemisorption
after the dissociation of the H$_2$ molecules, so that elucidating
the H$_2$ molecular physisorption stage that precedes the dissociation
is clearly very important.\cite{Costanzo} 

In view of the above considerations we have decided to investigate the
interaction of atoms and small molecules (X=Ar, CO, H$_2$, H$_2$O) 
with benzene and with the ideal planar graphene surface, by adopting 
the most recent theoretical approaches based on the Density Functional Theory
(DFT), explicitly developed to describe dispersion, van der Waals (vdW) 
interactions neglected\cite{Kohn} in standard DFT calculations, and thus 
being able to reproduce even weak physisorption processes.
In particular, the interactions of Ar and H$_2$ with benzene and graphene are 
representative of purely dispersion binding, while those involving CO and
H$_2$O are typical examples of mixed (dispersion/electrostatic and also
Hydrogen-bond for H$_2$O) interactions: in fact the polar water molecule
is characterized by a substantial dipole moment, while the weakly polar CO
molecule behaves as an electric quadrupole.\cite{Lakhlif}   

In the last few years several practical methods have been proposed
to make DFT calculations able to accurately describe vdW effects (for a
recent review, see, for instance, refs. \onlinecite{Riley,MRS,Klimes}).
We have developed a family of such methods, all based on the generation
of the Maximally Localized Wannier Functions (MLWFs),\cite{Marzari}
successfully applied to a variety of 
systems, including small molecules, water clusters, 
graphite and graphene, water layers interacting with graphite, 
interfacial water on semiconducting substrates,
hydrogenated carbon nanotubes, 
molecular solids, the interaction of rare gases and small molecules 
with metal surfaces,...
For the sake of clarity in the following we briefly summarize
the evolutionary process of these methods by reporting the improvements which
have been implemented during the time, together with the adopted nomenclature: 

DFT/vdW-WF denotes the original 
scheme\cite{silvprl,silvmetodo,silvsurf,CPL,silvinter,ambrosetti,Costanzo,Ar-Pb,Ambrosetti2012} 
where the vdW correction to the binding energy is obtained by using the basic 
information (center positions and spreads) given by the MLFWs and the
functional proposed by Andersson, Langreth, and Lundqvist (see eq. (10) of
ref. \onlinecite{Langreth96}) to evaluate the C$_6$ coefficients that 
characterize the long-range interaction between two separated fragments 
of matter. 
The DFT/vdW-WF method has been independently improved by Andrinopoulos
{\em et al.}\cite{Mostofi} to consider partly occupied Wannier
functions and $p$-like states.

The subsequent DFT/vdW-WF2 version\cite{C3} is based on the simpler
London expression and takes into account the intrafragment overlap
of the localized Wannier functions, leading to a considerable improvement
in the evaluation of the $C_6$ vdW coefficients.

DFT/vdW-WF2s\cite{PRB2013} indicates a modification of DFT/vdW-WF2 to take
metal-screening effects into account to be applied to the study of 
adsorption of rare gases and small molecules on metal surfaces.

Finally, the latest DFT/vdW-QHO-WF method\cite{QHO-WF} combines
the Quantum Harmonic Oscillator (QHO) model with the MLWFs, in such a
way to be no longer restricted to the case of well separated 
interacting fragments and to include higher than pairwise energy contributions,
coming from the dipole--dipole coupling among quantum oscillators.
In the specific case of adsorption on metal surfaces a long-range damping
factor has been introduced\cite{QHO-WF} to take metal-screening effects
into account.

Here we apply the DFT/vdW-QHO-WF approach to investigate 
atoms and small molecules weakly
interacting with benzene and with the ideal planar graphene surface.
Our results will be compared to the best available, reference experimental
and theoretical values, and to those obtained by other DFT
vdW-corrected schemes, including PBE+D,\cite{PBE+D} 
vdW-DF,\cite{Dion,Langreth07} vdW-DF2,\cite{Lee-bis} rVV10,\cite{Sabatini}
and by the simpler Local Density Approximation (LDA) and semilocal Generalized
Gradient Approximation (GGA, in the PBE flavor\cite{PBE}) approaches.
In the PBE+D scheme DFT calculations at the PBE level are corrected
by adding empirical $C_6/R^6$ potentials with parameters determined by fitting
accurate energies for a large molecular database, while in the 
vdW-DF, vdW-DF2, and rVV10 methods vdW effects are included by 
introducing DFT non-local correlation functionals.
While for the X-benzene systems both DFT/vdW-QHO-WF and the other considered 
vdW-corrected schemes perform well, it turns out that an accurate 
description of the
X-graphene interaction requires a proper treatment of many-body contributions
and of short-range screening effects, so that modified versions of the 
DFT/vdW-QHO-WF method are introduced and their performance is assessed.

\section{Method}
Here we briefly review the DFT/vdW-QHO-WF method; additional details can be
found in ref. \onlinecite{QHO-WF}.

For a system of $N$ three-dimensional QHOs the exact total energy can be 
obtained\cite{Cao,Donchev,Tkatchenko12,Reilly,QHO} 
by diagonalizing the $3N \times 3N$ 
matrix $C^{QHO}$, containing $N^2$ blocks $C_{ij}^{QHO}$ of size $3 \times 3$:

\begin{equation}
C_{ii}^{QHO} = \omega_i^2{\bf I}\,\,\,\, ; \,\,\,\,
C_{i\neq j}^{QHO} = \omega_i\omega_j{\sqrt {\alpha_i\alpha_j}}T_{ij}
\label{CQHO}
\end{equation}

where ${\bf I}$ is the identity matrix, $T_{ij}$ is the dipole-dipole
interaction tensor, and
$\omega_i$ and $\alpha_i$ are the characteristic frequency and
the static dipole polarizability, respectively, of the $i$-th oscillator.
The interaction (correlation) energy is given by the
difference between the ground state energy of the {\it coupled} system of
QHOs (proportional to the square root of the eigenvalues
{$\lambda_p$} of the $C^{QHO}$ matrix) and the ground state energy of the
{\it uncoupled} system of QHOs (derived from the characteristic frequencies):

\begin{equation}
E_{c,QHO} = 1/2 \sum_{p=1}^{3N} \sqrt{\lambda_p} - 3/2 \sum_{i=1}^{N}\omega_i
\,.
\label{interac}
\end{equation}

The so-computed interaction energy naturally includes many body
energy contributions, due to the dipole--dipole coupling among the QHOs;
moreover, it can be proved\cite{QHO} that 
the QHO model provides an efficient description of the
correlation energy for a set of localized fluctuating dipoles at an effective
Random Phase Approximation (RPA)-level. This is important because, 
differently from other schemes, RPA includes the effects of long-range
screening of the vdW interactions,\cite{Goltl} which are clearly 
of relevance, particularly for extended systems.\cite{Klimes,Bucko,JCTC}

The QHO model can be combined with the MLWF technique by 
assuming that each MLWF is represented by a three-dimensional isotropic
harmonic oscillator, so that
the system is described as an assembly of fluctuating dipoles.
By considering\cite{QHO} the Coulomb interaction 
between two spherical Gaussian charge densities
to account for orbital overlap at short distances (thus introducing
a short-range damping):

\begin{equation}
 V_{ij}=\frac{{\it erf}(r_{ij}/\sigma_{ij})} {r_{ij}}\;,
\label{Vij}
\end{equation}

where $r_{ij}$ is the distance between the $i$-th and the $j$-th 
Wannier Function Center (WFC), and  $\sigma_{ij}$ is an effective width,
$\sigma_{ij}=\sqrt{S_i^2 + S_j^2}$, where $S_i$ is the spread 
of the $i$-th MLWF.  
Then, in Eq. \eqref{CQHO} the dipole interaction tensor is\cite{QHO}

\begin{equation}
T_{ij}^{ab} = -\frac{3r_{ij}^ar_{ij}^b-r_{ij}^2\delta_{ab}} {r_{ij}^5}
\left( {\it erf}(\frac {r_{ij}} {\sigma_{ij}})-
\frac{2} {\sqrt{\pi}}\frac{r_{ij}} {\sigma_{ij}}
e^{-(\frac{r_{ij}} {\sigma_{ij}})^2} \right) +
\frac{4} {\sqrt{\pi}}\frac{1} {\sigma_{ij}^3}
\frac{r_{ij}^ar_{ij}^b} {r_{ij}^2}
e^{-(\frac{r_{ij}} {\sigma_{ij}})^2} 
\label{T}
\end{equation}

where $a$ and $b$ specify Cartesian coordinates ($x,y,z$), 
$r_{ij}^a$ and $r_{ij}^b$ are the respective components of the
distance $r_{ij}$, and $\delta_{ab}$ is the
Kronecker delta function.

Moreover, as in ref. \onlinecite{C3}, adopting a simple classical theory,
the polarizability
of an electronic shell of charge $eZ_i$ and mass $mZ_i$, tied to a heavy
undeformable ion, is written as
                                                                                
\begin{equation}
\alpha_i = \zeta \frac{Z_i e^2}{m\omega_i^2}\,.
\label{alfa}
\end{equation}

Then, given the direct relation between polarizability and 
volume,\cite{polvol} we assume that $\alpha_i = \gamma S_i^3$,
where $\gamma$ is a proportionality constant, so that the orbital volume is
expressed in terms of the $i$-th MLWF spread, $S_i$.

Similarly to ref. \onlinecite{QHO}, we combine 
the QHO model, which accurately describes the long-range
correlation energy, with a given semilocal, Generalized Gradient Approximation
(GGA) functional (PBE in
our case), which is expected to well reproduce short-range correlation
effects, by introducing an empirical parameter $\beta$ that
multiplies the QHO-QHO parameter $\sigma_{ij}$ in Eq. \eqref{Vij}.
The three parameters $\beta$, $\gamma$, and $\zeta$ are set up by minimizing the
mean absolute relative errors (MARE), measured with respect
to high-level, quantum-chemistry reference values relative to the
S22 database of intermolecular interactions,\cite{Jurecka} a widely
used benchmark database, consisting of weakly interacting molecules
(a set of 22 weakly interacting dimers mostly of biological importance), 
with reference binding energies calculated
by a number of different groups using high-level
quantum chemical methods. 
By taking PBE as the reference DFT functional, we get: 
$\beta=1.39$, $\gamma=0.88$, and $\zeta=1.30$.\cite{QHO-WF}    
Once the $\gamma$ and $\zeta$ parameters are set up, 
both the polarizability $\alpha_i$ and the characteristic frequency 
$\omega_i$ are obtained 
just in terms of the MLWF spreads (see Eq. (5) and below).

As anticipated above, the DFT/vdW-QHO-WF scheme can be improved to
achieve a better description of screening effects.
In fact the presence of the environment acts to screen the dipolar
fluctuations and both short- and long-range effects should be
properly included.
As a first choice we follow the strategy proposed in 
ref. \onlinecite{Tkatchenko12}, where the frequency-dependent 
polarizability tensor of finite-gap molecules and solids is obtained
by using the self-consistent screening (SCS) equation of classical 
electrodynamics:

\begin{equation}
\alpha_{ab}^{SCS}({\bf r},i\omega)=\alpha_{ab}({\bf r},i\omega)+
\sum_{cd} \alpha_{ac}({\bf r},i\omega) \int d{\bf r'}\,
U_{cd}({\bf r}-{\bf r'}) \alpha_{db}^{SCS}({\bf r'},i\omega) \; .
\end{equation}

Since our MLWFs are represented as a collection of isotropic QHOs
(differently from ref. \onlinecite{Tkatchenko12}, where the atoms
of the system were described as QHOs)
the above equation can be discretized as follows :

\begin{equation}
\alpha_{i,ab}^{SCS}=\alpha_{i,ab}+\sum_{cd} \alpha_{i,ac}
\sum_{j\neq i} U_{ij,cd}\, \alpha_{j,db}^{SCS} \;,
\end{equation}

where the tensor $U$ is related to the dipole interaction tensor defined 
above : $U_{ij,ab}=-T_{ij}^{ab}$.
                                                                                
Then, starting from an isotropic polarizability (see above) :

\begin{equation}
\alpha_{i,ab} = \alpha_i \delta_{ab} = \gamma S_i^3 \;,
\end{equation}

by focusing on the isotropic contribution of the SCS polarizability only, 
one obtains :
 
\begin{equation}
\alpha_i^{SCS} = 1/3 \sum_a \alpha_{i,aa}^{SCS} \;,
\end{equation}

where :

\begin{eqnarray*}
\alpha_{i,aa}^{SCS} & = &\alpha_i+ \alpha_i
\sum_{j\neq i} \sum_c U_{ij,ac} \alpha_{j,ca}^{SCS} = \\
 & = & \alpha_i+ \alpha_i
\sum_{j\neq i} U_{ij,aa} \alpha_{j,aa}^{SCS} +
\alpha_i
\sum_{j\neq i} \sum_{c\neq a} U_{ij,ac} \alpha_{j,ca}^{SCS} = \\
 & = & \alpha_i+ \alpha_i
\sum_{j\neq i} U_{ij,aa} \alpha_{j,aa}^{SCS} +
\mathcal{O}(U^2) \;.
\end{eqnarray*}

We have explicitly checked, that neglecting the $\mathcal{O}(U^2)$ terms
in the above expression, has negligible effects (less than 0.1 meV in the
binding energy for all the systems considered in the present study),
so that one can safely use the simpler, approximated expression:

\begin{equation}
\alpha_{i,aa}^{SCS}\simeq \alpha_i+ \alpha_i
\sum_{j\neq i} U_{ij,aa} \alpha_{j,aa}^{SCS} \;,
\end{equation}

representing a system of linear equations which
can be easily solved by a matrix inversion:

\begin{equation}
\alpha_{i,aa}^{SCS}=\sum_j (A_a^{-1})_{ij} \alpha_j \;,
\end{equation} 

where :
                                                                                
\begin{equation}
(A_a)_{ii}=1 \;\;, \;\;\; (A_a)_{ij}=-\alpha_i U_{ij,aa} \;\; (j\neq i)  \;.
\end{equation}

The isotropic, SCS polarizability $\alpha_i^{SCS}$ then replace the
original polarizability $\alpha_i$ in the matrix $C^{QHO}$ of Eq. (1).
We denote the version of our method modified by SCS effects as
DFT/vdW-QHO-WF-SCS.

Additional investigations\cite{QHO-SCS-SR} suggested that, 
within the QHO model, a better treatment of screening effects can be 
accomplished by range-separating the $U$ interaction tensor into a short
and a long-range component: the short-range (SR) component is used for the 
short-range SCS of the polarizabilities, while the long-range (LR) one is used 
in the QHO Hamiltonian. In this way no double counting of long-range 
screening effects is present,\cite{QHO-SCS-SR} which represents an advantage
with respect to the SCS method described above, 
especially in highly anisotropic systems, where the long-range effects tend 
to dominate over the short-range screening.
In practice, this new scheme is implemented as follows :

{\it (i)} the SCS procedure is only restricted to SR : 

\begin{equation}
\alpha_{i,aa}^{SCS}=\alpha_i+ \alpha_i
\sum_{j\neq i} U_{ij,aa}^{SR} \alpha_{j,aa}^{SCS} \;,
\end{equation}

where $U_{ij,aa}^{SR}=(1-f_{ij,aa}) U_{ij,aa}$, $f_{ij,aa}$ being a suitable
range-separating damping function;

{\it (ii)} in Eq. (1) the dipole-dipole interaction tensor is replaced
by its LR component  :

\begin{equation}
T_{ij}^{ab LR} = f_{ij,ab} T_{ij}^{ab} \;.
\end{equation}  

In ref. \onlinecite{QHO-SCS-SR} the range separation is enforced by
introducing a Fermi-type function, depending on an empirical parameter
to be fitted on minimizing the mean absolute relative error for
a reference database. 
In the present approach we have instead adopted an alternative 
expression for the damping function, namely :

\begin{equation}
f_{ij,ab} = U_{ij,ab}/D_{ij,ab} \;,
\end{equation}  

with the constraint that $f_{ij,ab}$ is set to zero if the above
expression leads to a negative (unphysical) value; here
$D_{ij,ab}$ indicates the $U$ interaction tensor at large
distances :  

\begin{equation}
D_{ij}^{ab} = \frac{3r_{ij}^ar_{ij}^b-r_{ij}^2\delta_{ab}} {r_{ij}^5}\; .
\label{D}
\end{equation}

This expression has the advantage that no additional empirical 
parameter is added. Moreover, extensive testing show that it is
more suited to our approach where the QHOs mimic the MLWFs instead
of the atoms as in ref. \onlinecite{QHO-SCS-SR}.
This new scheme, which is basically characterized by restricting the SCS 
procedure to short-range interactions only, will be referred to as 
DFT/vdW-QHO-WF-SCS-SR.

The calculations have been performed 
with both the CPMD\cite{CPMD} and the Quantum-ESPRESSO ab initio 
package\cite{ESPRESSO}
(in the latter case the MLWFs have been generated as a post-processing 
calculation using the WanT package\cite{WanT}). 
Electron-ion interactions were described using norm-conserving
pseudopotentials and the PBE reference DFT functional\cite{PBE}
which was adopted also in refs. \onlinecite{QHO,QHO-WF} and represents 
one of the most popular GGA choices.
In our calculations graphene was modeled using a supercell containing
72 C atoms (similarly to what assumed in previous theoretical 
studies\cite{Rubes,ambrosetti}), in such a way to prevent the X fragment
from interacting with its periodic images, and an empty region
of about 15 \AA\ width was left among the graphene replicas, 
in the direction orthogonal to the graphene plane. The 
in-plane geometry was fixed to the one determined experimentally
(C--C distance = 1.421 \AA).
The sampling of the Brillouin Zone was limited to
the $\Gamma$ point, again as done in previous studies.\cite{Rubes,ambrosetti}

In all the cases the X atom or molecule has been placed above the center
of the benzene ring or of a Carbon hexagon of graphene (in agreement with the
favored configurations suggested by previous studies) and
the distance between the X fragment and the benzene or graphene plane has
been optimized. In the case of X=H$_2$ and X=H$_2$O the 
molecule is orthogonal to the plane (with the H atoms pointing downwards
for water), while for X=CO the molecule is parallel, again in agreement
with previous findings (see references listed in the Result section and
in the related tables).

Although in principle the values of the three parameters 
$\beta$, $\gamma$, and $\zeta$ introduced above could be reoptimized
as the new DFT/vdW-QHO-WF-SCS and DFT/vdW-QHO-WF-SCS-SR schemes are applied,
we have maintained their original values (namely at the DFT/vdW-QHO-WF
level) both because in this way the effects of the SCS and SCS-SR
corrections can be more clearly assessed, and also because a 
reoptimization using the S22 database (consisting of interacting 
molecules) as the reference set would be 
of doubtful utility since here our primary interest is the application
to extended systems such as those involving graphene.  

\section{Results and Discussion}
In Tables I and II we report the binding energy and the equilibrium
distance for X-benzene and X-graphene systems, respectively.
These quantities have been computed using the DFT/vdW-QHO,
DFT/vdW-QHO-SCS, and DFT/vdW-QHO-SCS-SR methods described above, but also
a variety of other 
vdW-corrected schemes, including PBE+D, vdW-DF, vdW-DF2, rVV10,
and the simpler (non-vdW-corrected) LDA and semilocal GGA
(in the PBE flavor) approaches.

As can be seen looking at Table I, for the X-benzene systems
all the considered vdW-corrected methods perform reasonably well.
In more detail, with respect to the DFT/vdW-QHO scheme, DFT/vdW-QHO-SCS
reduces the binding energy and increases the equilibrium distance, while 
DFT/vdW-QHO-SCS-SR gives intermediate results.
On the whole DFT/vdW-QHO-SCS-SR clearly better reproduces the binding energy,
while the same is not evident for the equilibrium distance, also
considering that the reference values are affected by a significant 
uncertainty.

Concerning the other schemes, rVV10, whose performances are comparable
to those of DFT/vdW-QHO-SCS-SR, seems to give the best results.
As already pointed out in the literature,\cite{silvmetodo,Langreth07} the 
vdW-DF method, based on the revPBE GGA functional,\cite{revpbe} 
clearly overestimates the equilibrium distances; moreover, it turns out
that vdW-DF2 represents a significant improvement with respect to the
previous vdW-DF scheme, particularly concerning the equilibrium distances.
Not surprisingly the performances of the non-vdW-corrected LDA and
PBE methods are very poor, with LDA (PBE) which severely overestimates
(underestimates) the binding energies and underestimates (overestimates)
the equilibrium distances, thus indicating that for the systems we have
investigated a proper inclusion of vdW effects is crucial.

Coming to the calculations on X-graphene (see Table II),
the general trend is similar to that observed for 
X-benzene, however, important differences can be noticed.
In particular,
interesting quantities to look at, are represented by the ratios
between the value of the binding energy and of the equilibrium distance
relative to an X-graphene system and that of the corresponding 
X-benzene system (see Table III). One can observe that, for the
binding energy, using the LDA and PBE methods, the ratio is 
considerably lower than that between the best available, reference 
experimental and theoretical values (last two rows of Table III),
thus indicating that these non-vdW-corrected 
schemes are not able to properly describe long-range interactions
characterizing the weak binding between atoms/molecules and graphene.
Clearly, only adopting vdW-corrected methods one can recover these
effects. 

As far as the equilibrium distances are concerned, one can see that,
going from X-benzene to X-graphene systems, their behavior
is not simply correlated to the that of the corresponding
binding energies, namely
it is not necessarily true that a decrease in the binding energy leads
to an increase in the equilibrium distances; actually, most of the 
distances slightly decrease from X-benzene to X-graphene, 
but for X=H$_2$O where the equilibrium distance is almost unchanged.

As a way to assess the performances of the different schemes in a more
quantitative way, in Tables IV and V we report the  
Mean Absolute Relative Error of the computed binding energy, MARE$_e$, and 
of the equilibrium distance, MARE$_d$, together with their sum, MARE$_s$, 
by computing the error relative to the reference theoretical values
(when multiple reference data are available average values have been 
considered). The MARE$_e$ for both X-benzene and X-graphene systems is also 
shown in Fig. 1.
Of course these quantities can only be viewd as rough quantitative indicators,
also considering that the reference values are often quite scattered
(as, for instance, in the case of water interacting with graphene), 
nonetheless they convey the basic information.
To facilitate the performance assessment, in Tables IV and V, we have
listed the different methods in the order of decreasing MARE$_s$.
As can be expected, the pure-GGA PBE method performance is poor, as 
the result of a dramatic underbinding and strong overestimate of the
equilibrium distances.
Clearly the DFT/vdW-QHO-SCS-SR schemes performs well and represents an
evident improvement with respect to the previous DFT/vdW-QHO approach
(and also to the DFT/vdW-QHO-SCS variant). 
Interestingly, for the X-benzene systems DFT/vdW-QHO-SCS-SR is second 
only to rVV10, while for the X-graphene it gives the best results, thus
suggesting that it gives a better description of screening in extended
systems. In fact, vdW-DF, vdW-DF2, and rVV10 tend to overestimate 
the binding energy ratio (see Table III); this effect can be probably 
ascribed to the neglect of long-range screening effects by these 
methods.\cite{Klimes,Goltl} 
Therefore one can conclude that, while for the X-benzene systems all the 
considered vdW-corrected schemes perform reasonably well (the MARE$_e$
of PBE+D, vdW-DF2, rVV10, and DFT/vdW-QHO-SCS-SR is not larger
than 15\%), it turns out that an accurate description of the
X-graphene interaction requires a proper treatment of many-body contributions
and of both long- and short-range screening effects: 
in fact with PBE+D, vdW-DF2, 
and rVV10 the MARE$_e$ is larger than 17\%, and reduces to less then
10\% only with DFT/vdW-QHO-SCS-SR.
Quite interestingly, the worsening of the performances which characterize 
all the methods in going from X-benzene to X-graphene systems, is
much reduced (it is the smallest) with DFT/vdW-QHO-SCS-SR.   

The LDA performances deserve a separate comment, since, looking at Tables 
IV and V (see also Fig. 1), 
it appears that, while for X-benzene systems the LDA results
are very poor, for X-graphene they are almost comparable to those of
vdW-corrected schemes and are much better than those obtained by PBE.
This could lead to adopt a simple LDA approach in order to get a rough
description of graphene/graphite weakly interacting with small fragments.
However these relatively good LDA performances should not be overemphasized,
because they are the just the result of a cancellation of errors: in fact
LDA tends to overestimate the binding energy in small systems (see Tables
I and IV), that is at short and medium range 
(in particular the exchange contribution is
overestimated\cite{Berland}), while, in extended systems, it is not 
able to proper describe long-range correlation effects, which are
therefore dramatically underestimated.
Moreover, the common assumption\cite{Lin} that LDA always tends to overbinding
in systems where vdW interactions are important, so that
the LDA value provides an upper limit to the (absolute value of) the 
binding energy, can be misleading, as demonstrated (see Table II) by
the Ar-graphene case where LDA turns out to underestimate the 
interaction energy.  
In addition one must point out that with LDA the equilibrium distances are 
always significantly underestimated.
 
\section{Conclusions}
In conclusion, we have presented an improved version of the 
DFT/vdW-QHO-WF method, recently proposed to include the vdW  
interactions in DFT, that has been specifically developed to 
better describe short-range screening effects.
The new scheme has been applied to atoms and small molecules 
interacting with benzene and with the ideal planar graphene surface.
The computed binding energies and equilibrium distances have been
compared with those obtained by the original DFT/vdW-QHO-WF method and
by other vdW-corrected schemes, showing that the new 
DFT/vdW-QHO-SCS-SR approach represents a clear improvement with respect
to DFT/vdW-QHO-WF and, particularly in the description of X-graphene
systems, outperforms the other methods, thus indicating that in
extended systems a proper treatment of both many-body contributions
and of long- and short-range screening effects is essential.
Finally, the simple LDA approach, which performs much better than the 
semilocal PBE GGA, seems to offer a reasonable description of 
X-graphene systems, in line with previous observations,\cite{Berland}
although one should always be aware of the intrinsic limitations of the method.

\section{Acknowledgements}
We thank very much R. Sabatini for help in performing rVV10 calculations and
A. Ambrosetti for useful discussions.

\vfill
\eject

\begin{table}
\caption{Binding energy E$_b$ (in meV) and (in parenthesis) equilibrium 
distance R (in \AA) for X-benzene systems,
using different methods, compared with available
experimental and theoretical reference data.
R is defined as the separation between Ar, (the closest) H,
C, and O atoms, and the benzene plane, for X=Ar, X=H$_2$, X=CO, and X=H$_2$O,
respectively.}
\begin{center}
\begin{tabular}{|l|c|c|c|c|}
\hline
method            &   X=Ar   &  X=H$_2$ &   X=CO   & X=H$_2$O  \\ \tableline
\hline
LDA               & -74[3.24]&-101[2.40]&-153[3.09]&-222[3.03] \\        
PBE               & -11[3.90]& -21[2.94]& -25[3.68]& -84[3.46] \\     
\hline
PBE+D             & -50[3.50]& -58[2.57]& -77[3.19]&-172[3.18] \\    
vdW-DF            & -64[3.72]& -45[3.15]& -86[3.56]&-122[3.60] \\   
vdW-DF2           & -57[3.58]& -45[2.90]& -85[3.44]&-129[3.47] \\  
rVV10             & -51[3.51]& -46[2.72]& -85[3.23]&-147[3.31] \\
\hline
DFT/vdW-QHO       & -63[3.53]& -45[2.75]& -95[3.31]&-152[3.27] \\       
DFT/vdW-QHO-SCS   & -41[3.67]& -40[2.86]& -62[3.47]&-120[3.38] \\      
DFT/vdW-QHO-SCS-SR& -51[3.66]& -38[2.90]& -78[3.44]&-133[3.36] \\    
\hline
ref. expt.        & -51[3.50]$^a$&  --- & -76[3.24,3.44]$^{g,h}$&-141$\pm 12$[3.32$\pm 0.07$]$^{j,k,l}$ \\              
ref. theory       &-50$\leftrightarrow$-48[3.55]$^{b,c}$&-55$\leftrightarrow$-40[2.70]$^{d,e,f}$&-77[3.32]$^i$&-142$\pm 5$[3.35]$^{c,m,n,o}$ \\ 
\hline
\end{tabular}
\tablenotetext[1]{ref.\onlinecite{Brupbacher}.}     
\tablenotetext[2]{ref.\onlinecite{Koch}.}           
\tablenotetext[3]{ref.\onlinecite{Crittenden}.}     
\tablenotetext[4]{ref.\onlinecite{Donchev07}.}        
\tablenotetext[5]{ref.\onlinecite{Hubner}.}         
\tablenotetext[6]{ref.\onlinecite{Hamel}.}          
\tablenotetext[7]{ref.\onlinecite{Nowak}.}          
\tablenotetext[8]{ref.\onlinecite{Brupbacher93}.}   
\tablenotetext[9]{ref.\onlinecite{Nagy}.}           
\tablenotetext[10]{ref.\onlinecite{Stoicheff}.}     
\tablenotetext[11]{ref.\onlinecite{Gotch}.}         
\tablenotetext[12]{ref.\onlinecite{Cheng}.}         
\tablenotetext[13]{ref.\onlinecite{Zhao05}.}        
\tablenotetext[14]{ref.\onlinecite{Min}.}           
\tablenotetext[15]{ref.\onlinecite{Molnar}.}        
\end{center}
\label{table1}
\end{table}
\vfill
\eject

\begin{table}
\caption{Binding energy E$_b$ (in meV) and (in parenthesis) equilibrium 
distance R (in \AA) for X-graphene systems,
using different methods, compared with available
experimental and theoretical reference data.
R is defined as the separation between Ar, (the closest) H,
C, and O atoms, and the graphene plane, for X=Ar, X=H$_2$, X=CO, and X=H$_2$O,
respectively.
When reference data for X-graphene were not available, the 
corresponding values relative to X-graphite have been reported, which
probably slightly overestimate\cite{ambrosetti,Rubes} the binding energy with 
graphene, since graphite can be considered as an assembly of multiple
graphene layers.}
\begin{center}
\begin{tabular}{|l|c|c|c|c|}
\hline
method            &   X=Ar   &  X=H$_2$ &   X=CO   & X=H$_2$O  \\ \tableline
\hline
LDA               & -84[3.16]& -86[2.40]&-110[3.03]&-146[3.06] \\        
PBE               & -13[3.85]& -13[3.05]& -12[3.72]& -28[3.65] \\     
\hline
PBE+D             & -95[3.33]& -66[2.53]&-106[3.18]&-147[3.17] \\    
vdW-DF            &-137[3.54]& -77[2.98]&-160[3.47]&-140[3.64] \\   
vdW-DF2           &-113[3.41]& -67[2.80]&-144[3.29]&-129[3.36] \\  
rVV10             &-112[3.34]& -64[2.65]&-149[3.17]&-140[3.25] \\
\hline
DFT/vdW-QHO       &-136[3.34]& -50[2.75]&-147[3.21]&-137[3.26] \\       
DFT/vdW-QHO-SCS   & -88[3.45]& -41[2.86]& -85[3.35]& -91[3.37] \\      
DFT/vdW-QHO-SCS-SR&-112[3.49]& -40[2.86]&-117[3.36]&-113[3.38] \\    
\hline
ref. expt.        & -99$\pm 4$[3.0$\pm 0.1$]$^a$& -48[---]$^c$&-113$\pm 1$[---]$^e$&-156[---]$^h$ \\
ref. theory       &-116[3.33]$^b$& -56[2.58]$^d$&-120[3.02]$^f$,-112[3.0]$^g$&-140$\leftrightarrow$-72[3.26$\leftrightarrow$3.46]$^{g,i,k,l,m,n,o}$ \\
\hline
\end{tabular}
\tablenotetext[1]{ref.\onlinecite{Vidali} (on graphite).}        
\tablenotetext[2]{ref.\onlinecite{Tkatchenko06} (on graphite).}     
\tablenotetext[3]{ref.\onlinecite{Mattera} (on graphite).}     
\tablenotetext[4]{ref.\onlinecite{Rubes09}.}                   
\tablenotetext[5]{ref.\onlinecite{Piper} (on graphite).}         
\tablenotetext[6]{ref.\onlinecite{Zhang}.}            
\tablenotetext[7]{ref.\onlinecite{Lin}.}              
\tablenotetext[8]{ref.\onlinecite{Avgul} (on graphite).}         
\tablenotetext[9]{ref.\onlinecite{Rubes}.}         
\tablenotetext[10]{ref.\onlinecite{Jenness}.}         
\tablenotetext[11]{ref.\onlinecite{Kysilka}.}         
\tablenotetext[12]{ref.\onlinecite{Voloshina}.}         
\tablenotetext[13]{ref.\onlinecite{Ma}.}         
\tablenotetext[14]{ref.\onlinecite{Li}.}         
\tablenotetext[15]{ref.\onlinecite{Hamada}.}         
\end{center}
\label{table2}
\end{table}
\vfill
\eject

\begin{table}
\caption{Binding energy and equilibrium distance (in parenthesis) ratio 
between the value relative to an X-graphene
system and that of the corresponding X-benzene system. When multiple reference
data are available, ratios between average values have been considered.
The last column reports values averaged over the four X fragments.} 
\begin{center}
\begin{tabular}{|l|c|c|c|c|c|}
\hline
method            &   X=Ar   &  X=H$_2$ &   X=CO   & X=H$_2$O & average \\ \tableline
\hline
LDA               & 1.14[0.98]&0.85[1.00]&0.72[0.98]&0.66[1.01]&0.84[0.99] \\
PBE               & 1.18[0.99]&0.62[1.04]&0.48[1.01]&0.33[1.05]&0.65[1.02] \\ 
\hline
PBE+D             & 1.90[0.95]&1.14[0.98]&1.38[1.00]&0.85[1.00]&1.32[0.98] \\ 
vdW-DF            & 2.14[0.95]&1.71[0.95]&1.86[0.97]&1.15[1.01]&1.71[0.97] \\   
vdW-DF2           & 1.98[0.95]&1.49[0.97]&1.69[0.96]&1.00[0.97]&1.54[0.96] \\  
rVV10             & 2.30[0.95]&1.39[0.97]&1.75[0.98]&0.95[0.98]&1.60[0.97] \\
\hline
DFT/vdW-QHO       & 2.16[0.95]&1.11[1.00]&1.55[0.97]&0.90[1.00]&1.43[0.98] \\ 
DFT/vdW-QHO-SCS   & 2.15[0.94]&1.02[1.00]&1.37[0.97]&0.76[1.00]&1.32[0.98] \\
DFT/vdW-QHO-SCS-SR& 2.20[0.95]&1.05[0.99]&1.50[0.98]&0.85[1.01]&1.40[0.98] \\
\hline
ref. expt.        & 1.94[0.86]&--- [---] &1.49[---] &1.11[---]& --- \\
ref. theory       & 2.37[0.94]&1.18[0.96]&1.51[0.93]&0.75[1.00]&1.45[0.96] \\
\hline
\end{tabular}
\end{center}
\label{table3}
\end{table}
\vfill
\eject

\begin{table}
\caption{Mean absolute relative error of computed binding energy, MARE$_e$, and 
equilibrium distance, MARE$_d$, together with their sum, MARE$_s$, relative 
to X-benzene systems.}
\begin{center}
\begin{tabular}{|l|r|r|r||}
\hline
method            &MARE$_e$(\%)&MARE$_d$(\%)&MARE$_s$(\%) \\ \tableline
\hline
LDA               & 80.2   &  9.1   & 89.3 \\ 
PBE               & 60.3   &  8.2   & 68.5 \\
vdW-DF            & 15.2   &  9.0   & 24.2 \\
DFT/vdW-QHO-SCS   & 16.5   &  3.7   & 20.2 \\
DFT/vdW-QHO       & 15.8   &  1.3   & 17.1 \\
PBE+D             & 11.6   &  3.8   & 15.4 \\ 
vdW-DF2           & 10.0   &  3.9   & 13.9 \\ 
DFT/vdW-QHO-SCS-SR&  7.7   &  3.6   & 11.3 \\
rVV10             &  5.0   &  1.4   &  6.4 \\
\hline
\end{tabular}
\end{center}
\label{table4}
\end{table}
\vfill
\eject

\begin{table}
\caption{Mean absolute relative error of computed binding energy, MARE$_e$, and 
equilibrium distance, MARE$_d$, together with their sum, MARE$_s$, relative 
to X-graphene systems.}
\begin{center}
\begin{tabular}{|l|r|r|r||}
\hline
method            &MARE$_e$(\%)&MARE$_d$(\%)&MARE$_s$(\%) \\ \tableline
\hline
PBE               & 82.2   & 16.5   & 98.7 \\
vdW-DF            & 30.9   & 11.4   & 42.3 \\
LDA               & 31.0   &  5.4   & 36.4 \\ 
DFT/vdW-QHO-SCS   & 22.9   &  6.5  &  29.4\\
DFT/vdW-QHO       & 21.0   &  4.1   & 25.1 \\
PBE+D             & 20.8   &  3.3   & 24.1 \\ 
rVV10             & 20.0   &  2.9   & 22.9 \\
vdW-DF2           & 17.0   &  5.1   & 22.1 \\ 
DFT/vdW-QHO-SCS-SR&  9.9   &  7.0   & 16.9 \\
\hline
\end{tabular}
\end{center}
\label{table5}
\end{table}
\vfill
\eject

\begin{figure}
\centerline{
\includegraphics[width=17cm]{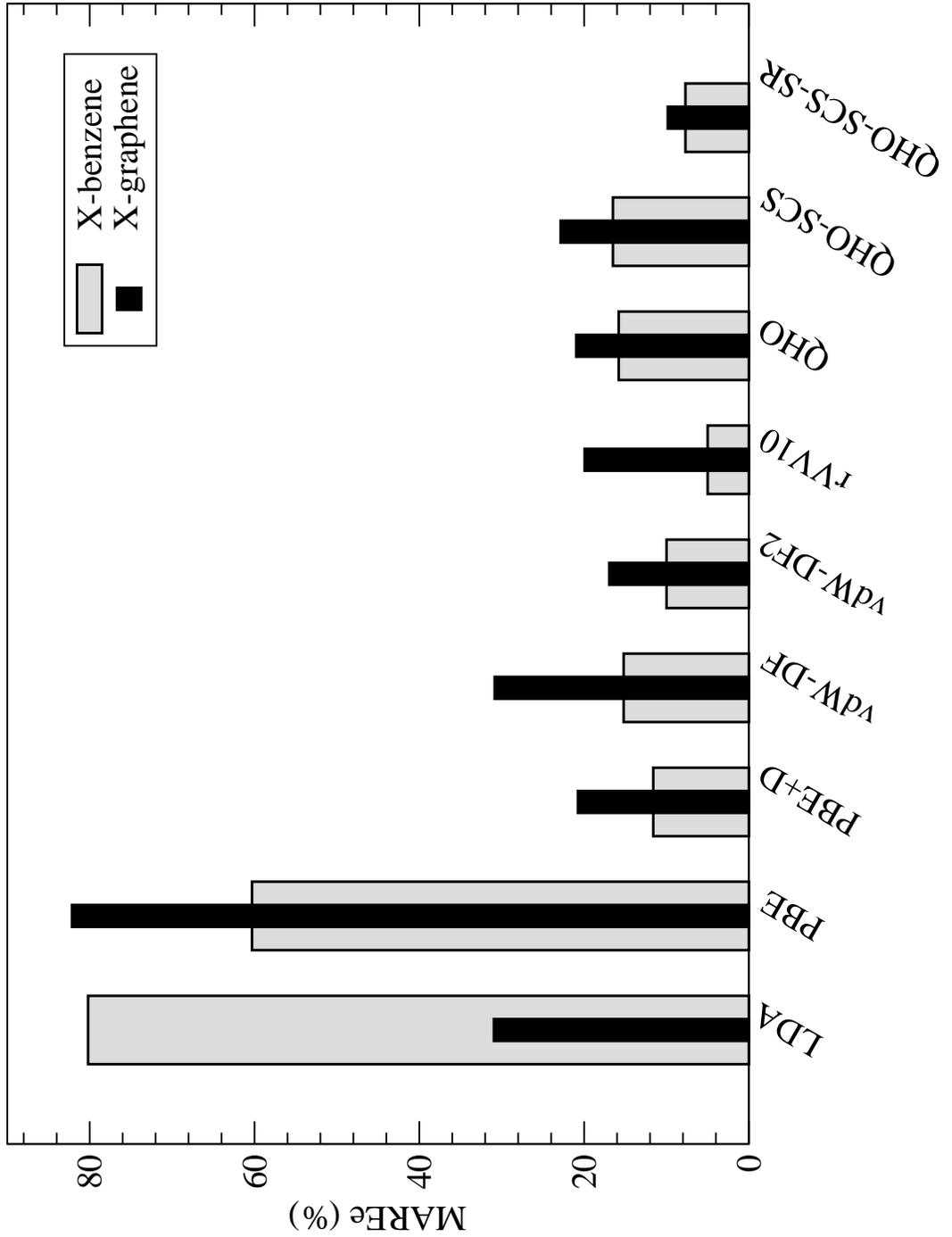}
}
\caption{MARE$_e$ for X-benzene and X-graphene using different methods.
         For brevity, QHO, QHO-SCS, and QHO-SCS-SR denote the DFT/vdW-QHO, 
         DFT/vdW-QHO-SCS, and DFT/vdW-QHO-SCS-SR methods, respectively.}
\label{fig}
\huge
\end{figure}

\end{document}